\documentclass{mem}
\usepackage{natbib}\usepackage{txfonts}\usepackage{balance}
\usepackage{graphicx}
\usepackage[a4paper]{hyperref}
\begin{document}

\title{Rotational Modulation of X-ray Emission from T Tauri Stars}

   \subtitle{}

\author{
S.G.\,Gregory\inst{1} 
\and M.\, Jardine\inst{1}
\and A.\, Collier Cameron\inst{1}
\and J.-F.\, Donati\inst{2}
        }

  \offprints{S. G. Gregory}

\institute{
University of St Andrews,
School of Physics and Astronomy, St Andrews,
KY16 9SS, United Kingdom
\and
Laboratoire d'Astrophysique, Observatoire Midi-Pyr\'en\'ees, 
14 Av. E. Belin, F-31400 Toulouse, France
\email{sg64@st-andrews.ac.uk}
}

\authorrunning{Gregory et al.}

\titlerunning{Rot. Mod. of X-ray Emission}

\abstract{
We have modeled the rotational modulation of X-ray emission from T Tauri stars assuming that
they have isothermal, magnetically confined coronae.  By extrapolating surface magnetograms we 
find that T Tauri coronae are compact and clumpy, such that rotational modulation arises from 
X-ray emitting regions being eclipsed as the star rotates.  Emitting regions are close to the 
stellar surface and inhomogeneously distributed about the star.  However some 
regions of the stellar surface, which contain wind bearing open field lines, 
are dark in X-rays.  From simulated X-ray light curves, obtained using stellar parameters from the 
Chandra Orion Ultradeep Project, we calculate X-ray periods and make comparisons with optically
determined rotation periods.  We find that X-ray periods are typically equal to, or are half of,
the optical periods.  Further, we find that X-ray periods are dependent upon the stellar inclination, 
but that the ratio of X-ray to optical period is independent of stellar mass and radius.
}

\maketitle{}

%-----------------------------------------------------------------

\section{Introduction}
One of the results from the Chandra Orion Ultradeep Project (COUP) was the 
detection of significant rotational modulation of X-ray emission from low 
mass pre-main sequence stars.  The detection of such modulation suggests that 
the coronae of T Tauri stars are compact and clumpy, with emitting regions 
that are inhomogeneously distributed across the stellar surface, and 
confined within magnetic structures that do not extend out to much beyond a 
stellar radius \citep{fla05}.  There is also evidence for much larger 
magnetic loops, possibly due to the interaction with surrounding circumstellar 
disks \citep{fav05,gia06}.  A model already exists for T Tauri coronae, where 
complex magnetic field structures contain X-ray emitting plasma close to the stellar 
surface, whilst larger magnetic loops and open field lines are able to carry 
accretion flows \citep{jar06,gre06a}. 

\begin{figure*}[t!]
\resizebox{\hsize}{!}{\includegraphics[clip=true]{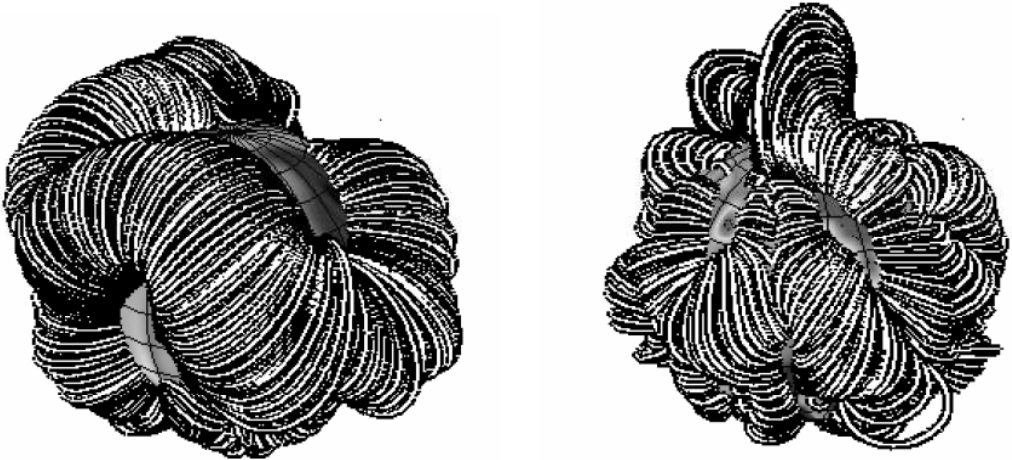}}
\caption{\footnotesize
Closed coronal structures showing field lines which contain X-ray emitting plasma, 
extrapolated from surface magnetograms of LQ Hya (left) and AB Dor (right).
The magnetic structures are compact and inhomogeneously distributed about the stellar
surface.
}
\label{extrap}
\end{figure*}

We use surface magnetograms derived from Zeeman-Doppler imaging to extrapolate the 
coronae of T Tauri stars (Fig. \ref{extrap}) using stellar parameters taken from the 
COUP dataset \citep{get05}.  The method for extrapolating the magnetic field is 
described by \citet{jar06} and \citet{gre06a}.  By considering isothermal coronae in 
hydrostatic equilibrium we have simulated X-ray light curves, for a range of stellar 
inclinations, and calculated X-ray periods using the Lomb Normalized Periodogram method 
\citep{gre06b}. We compare our results with those of \citet{fla05}, who demonstrate 
that those COUP stars which show clear evidence for rotationally modulated X-ray emission 
appear to have X-ray periods which are either equal to the optically determined 
rotation period $P_X = P_{opt}$ or are half of it $P_X = 0.5P_{opt}$.

%----------------------------------------------------------------

\section{X-ray and Optical Periods}
\begin{figure}[]
\resizebox{\hsize}{!}{\includegraphics[clip=true]{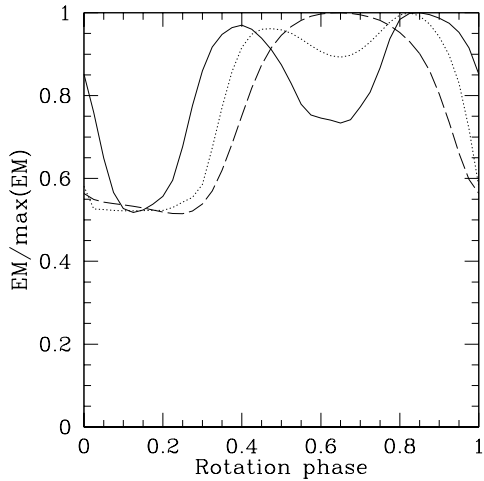}}
\caption{
\footnotesize
The variation in X-ray emission measure (EM) with rotation phase for the LQ Hya-like 
coronal structure shown in Fig. \ref{extrap}, for inclinations of 30$\,^{\circ}$ ({\it dashed}), 
60$\,^{\circ}$ ({\it dotted}) and 90$\,^{\circ}$ ({\it solid}).  There is clear rotational modulation 
of X-ray emission.
}
\label{lightcurve}
\end{figure}

\begin{figure}[]
\resizebox{\hsize}{!}{\includegraphics[clip=true]{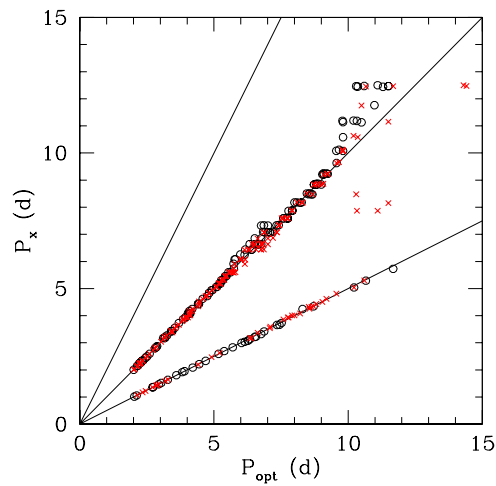}}
\caption{
\footnotesize
Comparison between our calculated X-ray periods and observed optical periods for the coronal 
fields in Fig. \ref{extrap} ({\em circles/crosses} for the LQ Hya/AB Dor-like coronal 
structures).  X-ray periods have been calculated for COUP stars from \citet{fla05} with 
randomly assigned inclinations.  Lines represent $P_X = [0.5, 1, 2]P_{opt}$.
}
\label{periods}
\end{figure}

We find that X-ray emitting regions are distributed across the stellar surface, but 
some regions remain dark in X-rays.  Emitting regions are close to the star and enter 
eclipse as it rotates. Clear rotational modulation of X-ray emission is apparent from 
plots of emission measure (EM) against rotational phase (Fig. \ref{lightcurve}).  The 
coronal magnetic field considered in Fig. \ref{extrap} (left panel) has two dominant 
emitting regions in opposite hemispheres.  \citet{fla05} argue that such a field 
structure would give rise to $P_X = 0.5P_{opt}$.  For large stellar inclinations we 
find that this is the case, but for small inclinations, our field structure with two 
dominant emitting regions in opposite hemispheres, gives rise to $P_X = P_{opt}$.  Thus 
the X-ray period depends on stellar inclination (Fig. \ref{periods}), but the ratio of 
X-ray to optical period is independent of stellar mass and radius \citep{gre06b}.  By 
selecting inclinations at random, and using different coronal magnetic field structures 
(Fig. \ref{periods}) we find that the amplitude of modulation strongly depends upon the 
stellar inclination and the distribution of X-ray emitting regions across the stellar 
surface.  In some cases an X-ray period of $0.5P_{opt}$ is found, but in the majority 
of cases, T Tauri coronae are so complex that it is difficult to disentangle the exact 
contribution to the X-ray light curve from any particular emitting region.  Therefore 
in the majority of cases we only recover X-ray periods which are equal to optical 
periods.

\begin{figure}[]
\resizebox{\hsize}{!}{\includegraphics[clip=true]{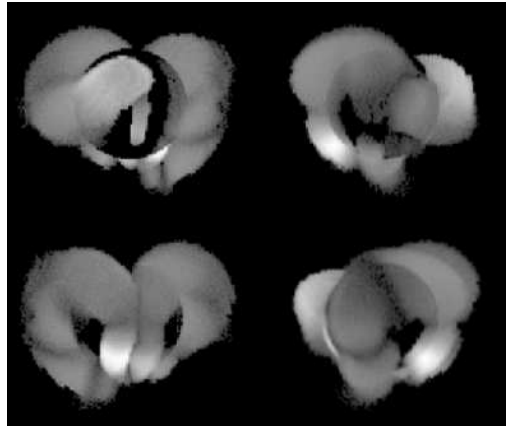}}
\caption{
\footnotesize
X-ray images corresponding to the LQ Hya field structure at an inclination of 90$\,^{\circ}$ - 
brighter regions indicate more X-ray emission. {\em Upper left} is for a rotational phase of 
0.1, where the brightest of the two dominant emitting regions is in eclipse, 
{\em upper right} 0.4, where both of the dominant emitting regions are visible, 
{\em lower left} 0.65 where the brightest emitting region is visible and 
{\em lower right} 0.85, where once again both of the dominant regions can be seen.
}
\label{lqhya}
\end{figure}

%----------------------------------------------------------------

\section{Discussion}
\citet{jar06} and \citet{gre06a} have shown that stars which have coronae that 
would naturally extend to beyond the corotation radius would have their outer corona 
stripped by the presence of a disk.  Any field line which passed through the disk at, or 
within the corotation radius was assumed to have the ability to carry an accretion flow 
and was therefore considered to be ``mass-loaded'' and set to be dark in X-rays.
We find that the influence of a disk makes no difference to the value of $P_X$ and very 
little difference to the amplitude of modulation.  Therefore the presence of a disk does 
not influence rotational modulation of X-ray emission, however, active accretion might 
and should be considered in future work.  

%----------------------------------------------------------------

\begin{acknowledgements}
SGG acknowledges the support from a PPARC studentship.  The authors thank Add van Ballegooijen 
who wrote the original version of the field extrapolation code and Keith Horne for valuable
discussions.
\end{acknowledgements}

\bibliographystyle{aa}

\end{document}